\documentclass[twocolumn]{aastex62}

\shorttitle{Firehose instability in multiple reconnection}
\shortauthors{Alexandrova et al.}

\begin{document}

\title{In situ evidence of firehose instability in multiple reconnection}

\correspondingauthor{Alexandra Alexandrova}
\email{alexandra.alexandrova@lpp.polytechnique.fr,\\ sasha.alexandrova@gmail.com}

 \author{ Alexandra Alexandrova }
\affiliation{Laboratoire de Physique des Plasmas, CNRS/Ecole Polytechnique/Sorbonne Universit\'{e}/Universit\'{e} Paris-Saclay/Observatoire de Paris, Palaiseau, France}
 
 \author{ Alessandro Retin\`{o} }
 \affiliation{Laboratoire de Physique des Plasmas, CNRS/Ecole Polytechnique/Sorbonne Universit\'{e}/Universit\'{e} Paris-Saclay/Observatoire de Paris, Palaiseau, France}
 
 \author{ Andrey Divin }
 \affiliation{Saint Petersburg State University, Saint Petersburg, Russia}
 
  \author{ Lorenzo Matteini }
 \affiliation{Department of Physics, Imperial College London, London SW7 2AZ, UK}
 
 \author{ Olivier Le Contel }
\affiliation{Laboratoire de Physique des Plasmas, CNRS/Ecole Polytechnique/Sorbonne Universit\'{e}/Universit\'{e} Paris-Saclay/Observatoire de Paris, Palaiseau, France}
 
 \author{ Hugo Breuillard }
 \affiliation{Laboratoire de Physique des Plasmas, CNRS/Ecole Polytechnique/Sorbonne Universit\'{e}/Universit\'{e} Paris-Saclay/Observatoire de Paris, Palaiseau, France}
  
 \author{ Filomena Catapano }
 \affiliation{Serco Itali, Department of Earth Observation, European Space Agency, ESRIN, Frascati ,Italy}
 
 \author{ Giulia Cozzani }
 \affiliation{Swedish Institute of Space Physics, Uppsala, Sweden}
 
 \author{ Ivan Zaitsev }
 \affiliation{Saint Petersburg State University, Saint Petersburg, Russia}
  
 \author{ Jan Deca }
 \affiliation{Laboratory for Atmospheric and Space Physics, University of Colorado, Boulder, USA}
 \affiliation{Institute for Modeling Plasma, Atmospheres and Cosmic Dust, NASA/SSERVI, USA}
 \affiliation{Laboratoire Atmosph\`eres, Milieux, Observations Spatiales (LATMOS), Universit\'e de Versailles \`a Saint Quentin, 78280 Guyancourt, France}
 
\begin{abstract}

Energy conversion via reconnecting current sheets is common in space and astrophysical plasmas. Frequently, current sheets disrupt at multiple reconnection sites, leading to the formation of plasmoid structures between sites, which might affect energy conversion. We present in situ evidence of the firehose instability in multiple reconnection in the Earth's magnetotail. The observed proton beams accelerated in the direction parallel to magnetic field and ion-scale fluctuations of whistler type imply the development of firehose instability between two active reconnection sites. The linear wave dispersion relation, estimated for the measured plasma parameters, indicates a positive growth rate of firehose-related electromagnetic fluctuations. Simulations of temporal evolution of the observed multiple reconnection by using a 2.5D implicit particle-in-cell code show that, as the plasmoid formed between two reconnection sites evolves, the plasma at its edge becomes anisotropic and overcomes the firehose marginal stability threshold, leading to the generation of magnetic field fluctuations. The combined results of observations and simulations suggest that the firehose instability, operating between reconnection sites, converts plasma kinetic energy into energy of magnetic field fluctuations, counteracting the conversion of magnetic energy into plasma energy occurring at reconnection sites. This suggests that magnetic energy conversion in multiple reconnection can be less efficient than in the case of the single-site reconnection.

\end{abstract}


\section{Introduction} 

The dynamics of magnetized, collisionless astrophysical plasmas implies the  
formation of current sheets accompanied by the accumulation of magnetic energy
\citep{parker1994}, and the fast release of such energy through magnetic reconnection \citep{priestforbes2000, yamada2010}. 
At the reconnection site, or X-line, 
nonlinear kinetic-scale processes mediate 
the large magnetohydrodynamical-scale rearrangement of the magnetic field, 
leading to plasma bulk acceleration, heating and non-thermal acceleration of particles.
Depending on the global conditions, the current sheet disruption may develop into single or multiple reconnection sites. During multiple reconnection, looped magnetic field structures (magnetic islands, plasmoids or flux ropes) tend to form between the adjacent X-lines. 
Though such plasmoid chains are considered to be an inevitable primary stage of the current sheet disruption \citep[][and references therein]{bhattacharjee2009,fulviapucci2018,uzdensky2016} as well as  an important stage of the single X-line evolution \citep{daughton2006}, the effect of plasmoid dynamics on the large-scale energy redistribution is not yet fully understood.

Numerical studies showed that plasmoid contraction
leads to {the acceleration} of trapped electrons \citep{drake2006} and ions \citep{drake2010} parallel {to the magnetic field}{(further referred to as parallel acceleration)} by first-order Fermi mechanism.
Accordingly, the multi-layered current sheets disrupted {into} plasmoids, can be responsible for non-thermal acceleration of particles, e.g., at the heliopause \citep{drake2010} and in the eruptive solar flares \citep{guidoni2016}.
Ion parallel acceleration is limited by the firehose instability, which manifests a{t} later stage{s} of plasmoids contraction in 2D  particle-in-cell \citep{drake2010} and 3D hybrid \citet{burgess2016} simulations.

In {space}{ plasmas}, statistical analysis of plasma properties showed that {the} firehose instability limits parallel acceleration of particles in the solar wind \citep{hellinger2006, bale2009}
and in magnetotail reconnection jets \citep{voros2011,wu2013}, except for an extreme case of long-duration {reconnection} exhaust \citep{hietala2015}, where {plasma} acceleration by reconnection appeared to {prevail over the effect of the} instability.  
The plasmoid dynamics in relation to the firehose instability, {however, was not studied } from observations.

In situ observations of multiple reconnection,
{showing a} plasmoid as well as neighboring reconnection sites, {were} provided in the Earth's magnetotail \citep{hwang2013, alexandrova2015}. 
A case study of {a} passage of two- X-lines by the spacecraft  \citep{alexandrova2016} revealed {the} highly variable magnetic field topology between the X-lines, representing complex stages of the plasmoid {evolution}.

{Here} we {present} in situ observations 
of multiple reconnection in the Earth's magnetotail \citep{alexandrova2016}, 
with a focus on the {ion} temperature anisotropy between {two} reconnection sites. 
The analysis of particle distribution {functions}, electromagnetic fluctuations and plasma stability conditions associated with the parallel temperature anisotropy of ions reveal typical conditions for the development of the firehose instability. 
The observations are supported by 2.5D implicit particle-in-cell (PIC) simulations {which allow to follow the} space-time dynamics of the plasmoid  in relation to {development of} the  instability.

\section{Overview of the multiple reconnection event in the Earth's magnetotail}

On 2002 August 18, between 17:07:00-17:13:00 UT, {the} Earth's magnetosphere was quiet.
Ground based observations of the ionosphere showed no signatures of a substorm. 
Cluster four-probe spacecraft \citep{escoubet2001}
was moving from northern to  southern hemisphere {across} the magnetotail current sheet, at about 17.7 Earth radii ($R_E$) tailward and 5 $R_E$ dawnward
in the Geocentric Solar Magnetic coordinate system (GSM).
The spacecraft detected {the} typical signature{s} of {a} consecutive passage of two reconnection sites and {of} the region in-between{,} where counterstreaming reconnection jets interact \citep[][]{alexandrova2016}. 
The signature{s} involve 
{three consecutive reversals in the following characteristics: (i) plasma bulk velocity; (ii) magnetic field component in the normal to the current sheet direction (reconnected field)  (iii) Hall magnetic field component}
(see Figure 1 in \cite{alexandrova2016}). 
A large value of the magnetic field {component} parallel to the current sheet ($\approx 16$ {nT}) and the jets' speed {($\approx 200$ km/s)} being smaller than the characteristic Alfv\'{e}n speed {($\approx 800$ km/s)} indicate that the structures were detected {at the ed{g}e} of the current sheet.
Between the reconnection sites, two Cluster probes observed two different stages of the jets{'} collision process, showing {the} formation and compression of a{n} ion{-}scale boundary separating {the} two counter-streaming jets, which is accompanied {by} strong wave activity (see Figures 2 and 3 in \cite{alexandrova2016}). {All t}his indicates that {the observed} plasmoid was not steady during Cluster observations, but {was} rather evolving in time.
{Here} we focus on the {region} between {the two} reconnection sites where {one of the Cluster probes, C1,} observed {enhanced} parallel temperature anisotropy of {ions}.

\section{In situ observations of firehose instability }

\begin{figure}[ht]
\centering  \includegraphics[width=0.5\linewidth]{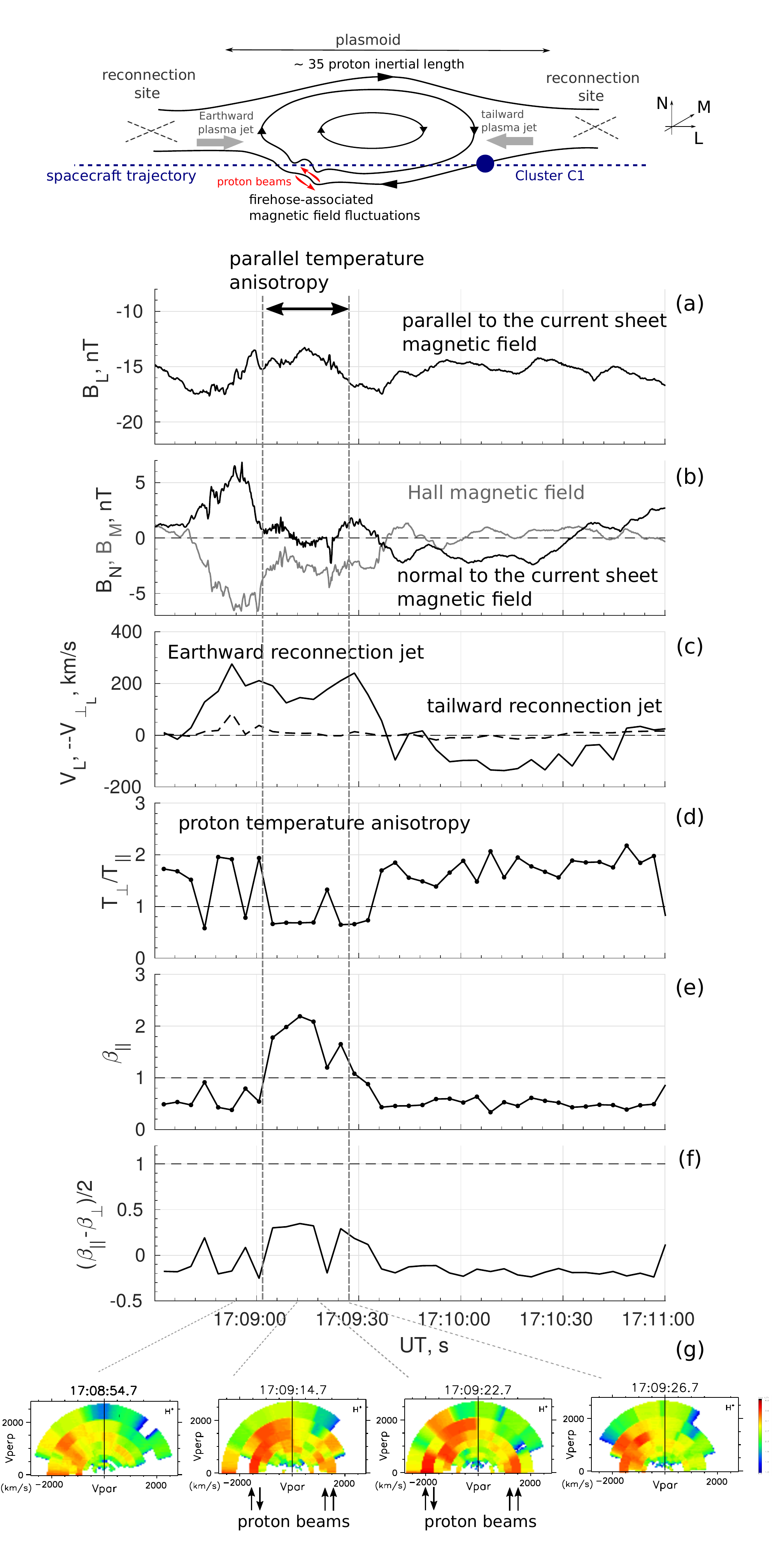}
\caption{Overview of Cluster C1 observations in the Earth's magnetotail on 2002 August 18 at 17:08:30-17:11:00 of (a) magnetic field components, (c) plasma bulk velocity in the parallel to the current sheet direction, $V_{L}$, (d) {ion } temperature ratio, (e) parallel plasma beta, (f) firehose instability threshold according to linear theory, (g) {ion} distribution functions for the selected times before, during and after the parallel temperature anisotropy observations. Sketch above the panels {represents} an interpretation of the observations.} 
\label{f.overview}
\end{figure}

Figure \ref{f.overview} presents Cluster C1 probe observations of
magnetic field ($22.4$ Hz resolution [Balogh et al., 2001]) 
and plasma ($H^{+}$ ion, $4$ s resolution plasma moments and distribution function{s} [Rème et al., 2001])  in the region {between the two reconnection sites}. {Measurements of $He^{+}$ and $O^{+}$ ions show densities of more than an order smaller than the $H^{+}$ {density}, thus we {analyze} $H^{+}$ ions {only}.} Data are represented in the current sheet  conventional coordinates LMN, where \textbf{L} is parallel to the current sheet and perpendicular to the reconnection line, \textbf{M} is parallel to the reconnection line, \textbf{N} is perpendicular to the current sheet. LMN was calculated {through} {M}inimum {V}ariance {A}nalysis {(MVA)} {of magnetic field} [Sonnerup and Cahill, 1967; Sonnerup et al., 2006] to the current sheet crossing prior to the reconnection activity at 16:40 - 17:00 UT for C1 probe \citep{alexandrova2016}. In GSM {coordinates}, $\mathbf{L} = (0.99, 0.03, −0.09)$, $\mathbf{M} = (−0.00, 0.95, 0.31)$, and $\mathbf{N} = (0.10, −0.31, 0.95)$. {In t}he time interval 17:08:30-17:11:00 {the} {spacecraft were located} below the magnetotail current sheet ($B_L<0$, Figure \ref{f.overview}a).
After the first X-line detection at $\sim$ 17:08:30, the probe C1 entered the earthward reconnection {jet} ($V_L>0$, $B_N>0$, $B_M<0$, Figure \ref{f.overview}a-c) and at $\sim$ 17:09:45, it detected the tailward {jet} ($V_L<0$, $B_N<0$, $B_M>0$, Figure \ref{f.overview}a-c) {coming} {from} the second X-line observed later at $\sim$ 17:11:00. 
The velocity perpendicular to the magnetic field, $V_{\perp_{L}}$, is negligible, supporting that the spacecraft is located at the current sheet edge where ions propagate mostly in the parallel direction.
The magnetic field configuration {is consistent with the observations of magnetic field and velocity, see} Figure \ref{f.overview} {sketch}. According to the {timing analysis applied to the reversals in the component of magnetic field normal to the current sheet, $B_N$,} \citep{alexandrova2015, alexandrova2016}, the plasmoid {observed in between the two reconnection sites} is moving tailward with {a} speed $U_L \approx 130$ km/s
{and} has a scale in the \textbf{L} direction of $\Delta L_o \approx 19500$ km $\approx 3$ $R_E \approx 3.4 $ $di$, where $di = 580$ km is the ion inertial length.
At the time 17:09:04-17:09:26, associated with the region between the first X-line and the jets{' collision site} {(marked by vertical lines in Figure \ref{f.overview})}, {C1 measurements showed parallel temperature anisotropy of {ions}, $T_{||}>T_\perp$}, where $T_{||}$ {and} $T_\perp$ are ion temperatures parallel {and perpendicular} to the local magnetic field, respectively (Figure \ref{f.overview}d). {The average temperature ratio is}
$T_\perp/T_{||} = 0.67$. 
According to linear analysis in the framework of {MHD theory}, the parallel temperature anisotropy {can} cause the firehose instability when
$\alpha = (\beta_\parallel - \beta_\perp)/2 > 1$, where $\beta_\parallel$ and $\beta_\perp$
are the ratios of the parallel and perpendicular plasma pressures, respectively, to the magnetic field pressure \citep{gary1998}.
At the time corresponding to the temperature anisotropy, the observations show an increase of $\beta_\parallel$ (Figure \ref{f.overview}e) as well as $\alpha$ (Figure \ref{f.overview}f). However the peak value is lower than the predicted firehose threshold.
{As for the other Cluster probes, C2 and C3 were far from the region of interest in M and in N direction, respectively, {therefore they are not used in this analysis}. The probe C4 was observing the region $\sim 10$ s later and showed parallel temperature anisotropy only in one measurement point, at 09:14.} 
                                                         
{During the time of interest $\approx 20$ s} associated with the ion parallel temperature anisotropy {({in }between the vertical lines in Figure \ref{f.overview})}, the {ion velocity distributions} show {the} presence of parallel and anti-parallel beams superposed to the bulk velocity of earthward propagating {ions} (anti-parallel to the background field), {see} Figure \ref{f.overview}g, 17:09:14.7 and 17:09:22.7 {times}. These beams are not observed {either} before ({Figure \ref{f.overview}g, time} 17:08:54.7) {or} after ({Figure \ref{f.overview}g, time} 17:09:26.7) the region of anisotropy. 

\begin{figure}[ht]
\centering  \includegraphics[width=0.6\linewidth]{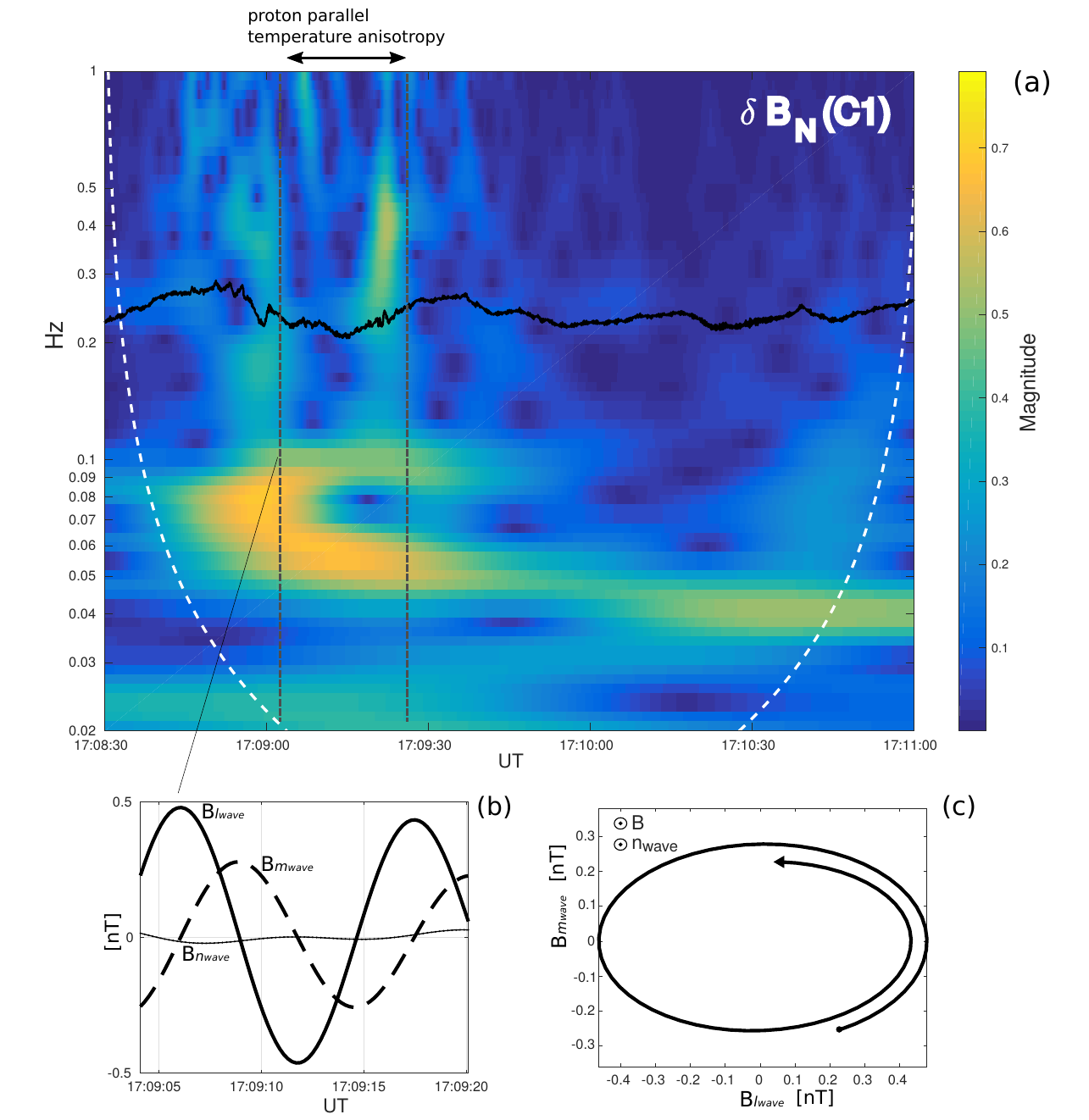}
\caption{(a) wavelet spectrum of the normal to the current sheet magnetic field component $B_N$, (b) waveform of the filtered fluctuations at $0.08-0.11$ Hz, (c) hodograph of the filtered fluctuations in the plane perpendicular to the wave normal.}
 \label{f.waves}
\end{figure}

We study {the} magnetic field fluctuations associated with the {observed} parallel temperature anisotropy.
Figure \ref{f.waves}a {shows} the wavelet spectrum of the magnetic field $B_N$ component, which is the reconnected magnetic field {component}.
The fluctuations which directly correspond to the time associated with the ion temperature anisotropy {and the $\beta_\parallel$ increase}  (17:09:04-17:09:20), 
are seen in the frequency range of $f_0 \approx 0.08-0.11$ Hz.
These frequencies are nearly twice lower than the corresponding {ion} cyclotron frequency $f_{ci}=0.22$ Hz.  
We apply a {bandpass} filter for the observed fluctuations
(17:09:04-17:09:20, $f = 0.08-0.11$ Hz ) and the 
Minimum Variance Analysis (MVA) \citep{sonnerupcahill1967} to calculate the
direction of propagation and the polarization of these
fluctuations \citep{thorne1973, smithtsurutani1976}. 
The wave components in the MVA coordinate system ($lmn_{wave}$) are {shown} in Figure \ref{f.waves}b.
The orientation {of the wave coordinate system }in  {the }LMN system is 
$\mathbf{l_{wave}} = (0.02, 0.64, 0.77)$, $\mathbf{m_{wave}} = (0.74, -0.53, 0.42)$, $\mathbf{n_{wave}} = (0.68, 0.56, 0.49)$.
The medium to minimum eigenvalue ratio is $188$, which indicates that the normal direction is well defined.
The ellipticity, defined as the square root of the medium to maximum eigenvalue ratio is $e=0.57$.
{With respect to} the background magnetic field, calculated as an average over the spacecraft spin, $\mathbf{B_{db}}$, 
the wave is propagating with an angle $\Theta = \angle (\mathbf{n_{wave}}, \mathbf{B_{bg}}) \sim 23^{o}$.
The MVA analysis contains a $180^{o}$ ambiguity in the normal vector $n_{wave}$ direction.
Under {the} assumption of wave propagation preferentially parallel to the background field, ${n_{wave_{L}}>0}$, the elliptically polarized wave exhibit right-hand rotation around the magnetic field,
see the wave hodograph in Figure \ref{f.waves}c.
The observed waves characteristics and frequency ranges
{are consistent with} the low branch whistler waves,
which are related to the linear firehose instability \citep{gary1998}.
According to the plasmoid tailward speed $U_L \approx -130$ km/s \citep{alexandrova2016}, the ion bulk speed $V_L \approx 160$ km/s (Figure \ref{f.overview}c between the dashed lines),   
the Alfv\'{e}n speed  $V_A\approx 800$ km/s, and the period of the ion gyration $\tau_i \approx 4$ s,  
we can roughly estimate the Doppler shift to be about $f_D = (V-U_L)/V_A T_i \approx 0.09 $ Hz. As long as background plasma propagates in the {direction }opposite to the wave {propagation }direction, the expected real wave frequency is {therefore} about $f \approx 0.2$ Hz, {which} is almost equal to the ion cyclotron frequency.  
The time interval of $16$ s corresponding to the temperature anisotropy {and $\beta_\parallel$ increase} is associated with one wave period. According to the plasmoid speed, the wave has a scale of $\Delta L \approx 1950$ km $\approx 0.3$ $R_E \approx 3.4$ $d_i$.
The amplitude of the observed fluctuation is $\delta \mathbf{|B|}/ |B_{bg}| \approx 0.03$.

\begin{figure}[ht]
\centering  \includegraphics[width=1\linewidth]{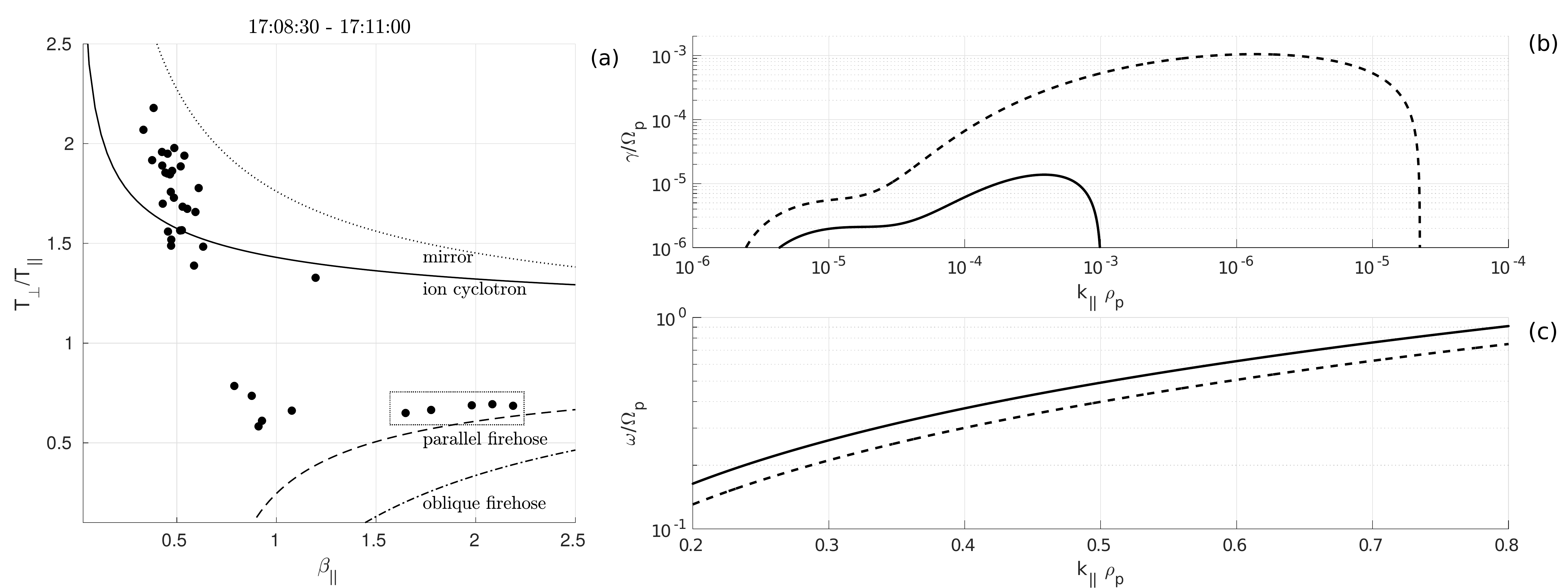}
\caption{(a) parallel temperature anisotropy of protons in relation to the parallel plasma beta in comparison to the marginal stability thresholds of the plasma instabilities, (b) instability growth rate and (c) real frequency of the corresponding fluctuations according to the stability analysis by WHAMP solver for the observed plasma parameters (solid lines) and presumed plasma parameters within the $30\%$ of inaccuracy of measurements (dashed lines).}
 \label{f.instability}
\end{figure}

In order to verify {whether }the firehose instability {was operating},  we perform {a} plasma stability analysis. Figure \ref{f.instability}a shows comparison between  $T_\perp/T_{||}$ and $\beta_\parallel$,
measured in {between the two reconnection sites} (17:08:30-17:11:00),
with the predictions of Vlasov linear theory for the
marginal stability thresholds of typical plasma instabilities 
(mirror, ion cyclotron, oblique and parallel firehose)
calculated for the maximum growth rate $\gamma\approx 10^{-3}${,} 
according to the fitting parameters from  \citet[]{hellinger2006}, Section 2 and Table 1.
The five data points, related to the parallel temperature anisotropy 
observed at 17:09:04-17:09:20, lie close to the parallel firehose threshold 
(marked with rectangle in Figure \ref{f.instability}a).
To investigate the growth rate of the possible firehose
instability, we solve the Vlasov-Maxwell equations by
using WHAMP solver \citep{roennmark1982},
assuming  {a} bi-Maxwellian gyrotropic {ion} distribution and different values of the wave vector direction.
It should be noted that the observed {ion} distribution function is neither bi-Maxwellian, nor gyrotropic.
In such a case a {recent } method developed in \citet{astfalkjenko2017} 
might give more correct estimates,
however it requires better resolution than Cluster provides.
Therefore, we use WHAMP calculations to obtain {indicative} estimates.
{T}he solution with positive growth rate is presented 
in Figure \ref{f.instability}b, solid line. 
For the observed plasma parameters:
the average magnetic field $\langle B\rangle = 15$ nT,
electron temperature $\langle T_e\rangle = 1$ keV,
{ion} density $\langle n\rangle = 0.16$,
{ion} temperature $\langle T_{\parallel} \rangle = 5$ keV,
and temperature anisotropy $ \langle T_{\perp}/T_{\parallel} \rangle = 0.67$.
The maximum growth rate is about $\gamma = 10^{-5}$ $\Omega_p$, 
where $\Omega_p$ is the {ion} frequency.
The dashed line in Figure \ref{f.instability}c represents the growth rate
calculated for {an} {ion} temperature enhanced by $30\%$ and anisotropy enhanced by $10\%$,
{to include} possible temperature underestimation 
{due to} instrument{al errors} \citep[Figure 6.7 of ][]{paschmanndaly2000}.
For the enhanced parameters, the growth rate reaches the marginal stability threshold of $\gamma = 10^{-3}$ $\Omega_p$.
For the observed parameters, 
the solution with the maximum positive growth rate corresponds to {a} wave with frequency $\omega = 0.48$ $\Omega_p$, which is about $0.07$ Hz, see Figure \ref{f.instability}c, 
propagating parallel to magnetic field and have right-hand polarization. The resulting electromagnetic fluctuations {are consistent} with the firehose instability.

\section{Reconstruction of observations with 2.5D PIC simulations}

\begin{figure}[ht]
\centering  \includegraphics[width=0.7\linewidth]{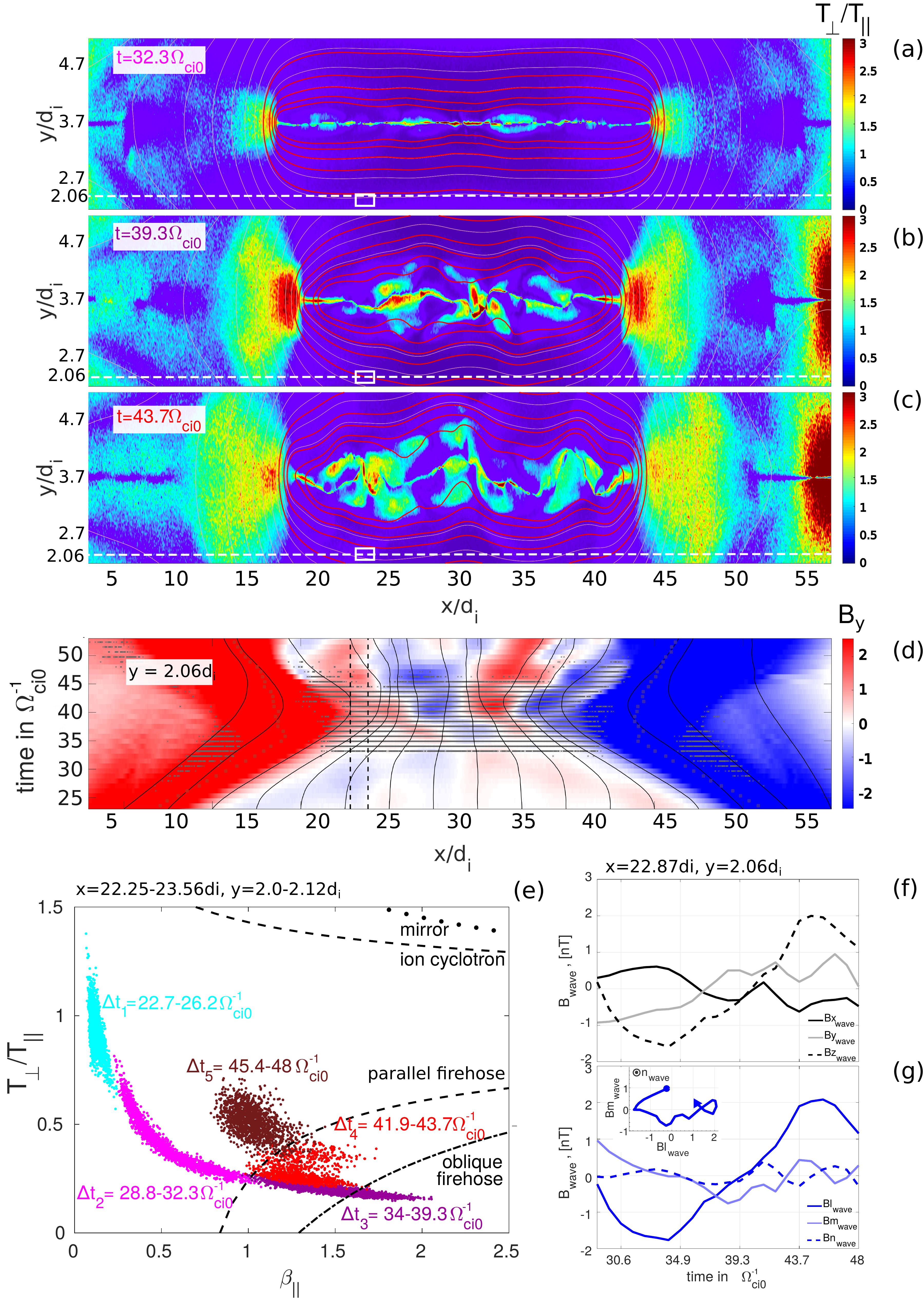}
\caption{$2.5$D iPIC simulations representing the dynamics of plasmoid forming between two reconnection sites. (a)-(c) parallel temperature anisotropy of {ions} (in color) and magnetic field configuration (red lines) for three stages of plasmoid development, (d) temporal evolution of the normal to the current sheet magnetic field (in color) and the regions where temperature anisotropy overcomes the marginal stability threshold of the firehose instability (gray crosses) for a cut in $y=2.06 d_i$. (e) temporal evolution of the temperature anisotropy in the selected region (white rectangle in panels (a)-(c)) in comparison to the marginal stability thresholds of plasma instabilities, (f) magnetic field fluctuations in the center of the selected region, (g) magnetic field fluctuations  in the minimum variance coordinate system and the hodograph in the plane perpendicular to the wave normal direction.}
 \label{f.3dpic}
\end{figure}

The {presented} {Cluster} magnetotail observations represent {only} single-spacecraft measurements of anisotropic plasma in {the} localized region of plasmoid between two X-lines, {leaving a} detailed investigation of the {overall} large-scale plasmoid {spatio-temporal evolution unresolved}.
Thus, we employ a 2.5D numerical simulation which reproduces the formation of a plasmoid, followed by its compression by reconnected plasma flows and {by the} development of firehose instability at later stages (Figure \ref{f.3dpic}a-\ref{f.3dpic}c). The iPIC3D implicit PIC code \citep{markidis2010} is used. The system of coordinates is as follows: the $\mathbf{x}$ axis is directed parallel to the reconnecting magnetic field {(corresponds to the \textbf{L} direction)}; the $\mathbf{y}$ axis is normal to the current sheet at time t=0 {(corresponds to the \textbf{N} direction)}; the $\mathbf{z}$ axis complements the right-hand triple {(corresponds to the \textbf{M} direction)}. The simulation is performed in 2D rectangular domain with the dimensions $(L_x,L_y)$. The model is translationally invariant in {the} third {direction, \textbf{z}, which is the direction of current}. The simulation is initialized with a pair of conventional Harris current sheets located at $y=L_y/4$ (active) and $y=3L_y/4$ (remains quiet in the present study, not shown). A uniform background population of density $n_b=0.1 n_0$ is added, with $n_0$ being the peak density of the Harris current sheet.
{The length unit (${d_i \, \approx \, 509}$ km) is computed {from} the plasmoid edge density (${\approx 0.35 n_0 = 0.2 }$ cm$^{-3}$) to ease comparison with observations. For such a normalization, the computational box dimensions amount to ${(L_x,L_y)=(60d_i,15d_i)}$, and the number of grid points in each dimension is ${(N_x,N_y)=(2304, 576)}$. The magnetic field $B_0$ (asymptotic magnetic field outside of {the} Harris sheets) is normalized to  $16$ nT, which is the largest magnetic field observed by the {Cluster} spacecraft {in} the parallel temperature anisotropy region (see Figure \ref{f.overview}a).}
Derived units are the Alfv\'en speed of $ 780 $ km/s and the ion cyclotron frequency $\Omega_{ci0} \sim 1.5$ $s^{-1}$.
The initial ion-to-electron temperature ratio is similar to that {of the  {Cluster} observations described above} ($T_i/T_e=5$). The ion-to-electron mass ratio is $m_i/m_e=256$, the ratio of the speed of light to the characteristic Alfv\'{e}n speed is 256.  

A localized X-point perturbation {\citep[see, e.g.,][]{divin2012}} ignites reconnection at $(0, L_y/4)$. Ion jets are formed at early stages once reconnection at the main X-line has reached the steady state. These jets propagate nearly unperturbed in the $\mathbf{x}$ and $-\mathbf{x}$ directions up to $t \approx 30 $ $\Omega_{ci0}^{-1}$.
To mimic the dynamical stage of the plasmoid evolution, we impose periodic boundary conditions and allow plasma jets to run head-to-head producing the domain-large plasmoid (Figures \ref{f.3dpic}a-\ref{f.3dpic}c). 
In essence, such periodic configuration can be viewed as interaction of two X-lines.
As reconnection progresses, ion flows compress the plasmoid, producing regions with different kinds of anisotropy: 
(i) perpendicular anisotropy (${T_{\perp}/T_{||}>1}$) is found in the plasmoid core and at reconnection fronts, ${x \approx 15.6 d_i}$ and ${x \approx 44 d_i}$ for the discussed times); 
(ii) parallel anisotropy (${T_{\perp}/T_{||}<1}$) is found typically in the low-$\beta$ regions {close to the edges of the plasmoid} (shown with deep blue color in Figures \ref{f.3dpic}a-\ref{f.3dpic}c).
In Figure \ref{f.3dpic}a-\ref{f.3dpic}c, we highlight a few field lines of constant magnetic flux to trace their evolution in time. At the beginning of strong interaction between the counterstreaming jets, {approximately at} ${t=30.6} \Omega_{ci0}^{-1}$, the plasmoid scale is $32 d_i \approx 16000$ km. This estimate is rather close to the scale of the {observed} magnetotail plasmoid described above.
In course of {simulation} time, the plasmoid shrinks and becomes about ${24} d_i \approx 12200$ km by ${t=36.7} \Omega_{ci0}^{-1}$.
{Further shrinking of the plasmoid, proceeding between ${t=36.7} \Omega_{ci0}^{-1}$ and ${t=45.5} \Omega_{ci0}^{-1}$  leads to a strong bending of the magnetic field lines {at the edges} of the plasmoid, while inside the plasmoid the magnetic field structure becomes strongly inhomogeneous following the nonlinear pressure growth.} 

In Figure \ref{f.3dpic}d we visualize the growth of magnetic field fluctuations at the {edge} of the plasmoid by plotting the time stack plots of $B_y$ component along a cut through ${y=2.06} d_i$. Very weak $B_y$ at earlier times ($t<30 \Omega_{ci0}^{-1}$, ${15 d_i < x < 45 d_i}$) are most likely attributed to the weak tearing instability present at earlier stage {\citep[see, e.g.,][Section 3]{pritchett1991}}.
Colliding jet fronts (peak $B_y$ locations are marked with the squares of corresponding colors) host large $B_y$ variations at 
$10 d_i<x<20 d_i$ and $40 d_i<x<50 d_i$. 
Note that the colorbar is compressed to reveal weaker $B_y$ perturbations in $20 d_i<x<40 d_i$.
Compression of the plasmoid produces conditions favorable for the excitation of the firehose instability.
Gray crosses indicate regions where the temperature anisotropy overcomes the firehose marginal stability threshold, 
calculated according to \citet{hellinger2006}.
Notably, fluctuations are strongly amplified at these times 
($30 \Omega_{ci0}^{-1}<t<40 \Omega_{ci0}^{-1}$),
before saturating at the de-compression stage.
Black lines in Figure \ref{f.3dpic}d are virtual streamlines of fluid elements 
located at $y=2.06 d_i$ (presuming that the velocity component along the $\mathbf{y}$ direction is unimportant).
The streamlines trace well the jet front locations. 
Although not exactly, the $B_y$ fluctuations follow the streamlines as expected for the firehose instability.

We focus on the region, marked by the white rectangle in Figures \ref{f.3dpic}a-\ref{f.3dpic}c: $22.25 d_i <x<23.56 d_i$ and ${2.0 d_i <y<2.12 d_i}$, which closely reproduces the spacecraft observations by a combination of field and plasma parameters. 
Figure \ref{f.3dpic}{d} shows temporal changes in the plasma
distribution of $T_{\perp}/T_{||}$ plotted against $\beta_{||}$ for the selected region. The thresholds for the plasma marginal stability are shown according to \citet{hellinger2006}. 
The ion temperature anisotropy in relation to $\beta_{||}$ indicate isotropic and stable plasma at the early simulation time ($\Delta t_1$, cyan). {A}t later stages{, the} plasma becomes more anisotropic and {the anisotropy} exceeds the parallel firehose threshold ($\Delta t_2$, pink and $\Delta t_3$, dark pink). {Then,} at the time corresponding to the active jets collision ($\Delta t_4$, red){, the} plasma becomes more isotropic and close to the marginal stability threshold, and at the end of simulations ($\Delta t_5$, dark red){, the} plasma becomes stable. 
{{In order to get a clear visualization}
in Figure \ref{f.3dpic}d, {we plot selected points which represent the overall behavior.}}
Figures \ref{f.3dpic}a, \ref{f.3dpic}b and \ref{f.3dpic}c reflect  plasmoid development stages for three selected times from the periods $\Delta t_2$, $\Delta t_3$ and $\Delta t_4$, respectively. 
The magnetic field temporal changes taken in the middle of the selected region, ${{x=22.9} ~d_i}$, ${{y=2.06} ~ d_i}$ represent one wave period (Figure \ref{f.3dpic}f). In two locations in $\mathbf{x}$ direction separated by about ${1.2 ~ d_i \approx 600}$ km the fluctuations are almost similar and shifted in time by $0.2$ $\Omega_{ci0}^{-1}$ $\approx 0.13$ s, indicating the wave phase speed {to be} about ${500} $ km/s, which is {smaller {than}} the corresponding Alfv\'{e}n speed (${780} $ km/s). The MVA analysis \citep{sonnerupcahill1967} gives the orientation of the normal $\mathbf{n_{wave}}=(0.92,0.11,0.3)$, the maximum and medium variance components are $\mathbf{l_{wave}}=(-0.25, 0.9, 0.34)$ and $\mathbf{m_{wave}}=(0.28,0.41,-0.87)$, respectively, in the PIC $(x,y,z)$ coordinates. 
Figure \ref{f.3dpic}e shows the wave in the $lmn_{wave}$ system and the rotation of magnetic field in the plane perpendicular to the normal direction, which indicate right-hand elliptical polarization.
Fluctuations show characteristics typical for the firehose instability, with the magnitude of about $\delta \mathbf{|B|} / B \approx 0.15$. 
Note, that the amplitude of the magnetic field fluctuations changes in time revealing the nonlinear evolution.

\section{Discussion}

At the {edge} of the plasmoid forming between two X-lines and having a scale of approximately $35$ ion inertial length{s} (about three Earth's radii), {Cluster} observations show distinctive signatures of the firehose instability, including 

(1) {a} parallel temperature anisotropy of {ions} of about $T_{\perp}/T_{\parallel}{\approx}$ $0.7$ (Figures \ref{f.overview}d,e and \ref{f.instability}a); 

(2) parallel and anti-parallel ion beams, superimposed to the bulk motion of plasma in the reconnection jet (Figure \ref{f.overview}g); 

(3) magnetic field fluctuations {at frequencies around} the ion cyclotron {frequency}, right-hand polarized and quasi-parallel propagating (Figure \ref{f.waves}); 

(4) plasma condition{s} corresponding to the marginal firehose state (Figure \ref{f.instability}a), 
with positive growth rate of {the }instability (Figure \ref{f.instability}b) for the firehose-like fluctuations (Figure \ref{f.instability}c) {in} approximately {the} ion cyclotron frequency range. 

2.5D iPIC simulations of the plasmoid formation between two reconnection jets running head-to-head (see Figures \ref{f.3dpic}a-\ref{f.3dpic}c){,} {with} plasma and magnetic field parameters, {similar to the ones observed,} show {that}

(5) in course of time the plasmoid is compressed{,} producing regions at {its} {edge} with {a} large parallel temperature anisotropy of the maximum value $T_{\perp}/T_{||} \sim 0.3$ (Figure \ref{f.3dpic}d, \ref{f.3dpic}e) 

(6) firehose-like electromagnetic fluctuations with relatively large amplitude $\delta B / B \approx 0.15$ (Figure \ref{f.3dpic}f, \ref{f.3dpic}g) arise when the {maximum} temperature anisotropy is reached (Figure \ref{f.3dpic}d, $t>30$ ${\Omega_{ci0}^{-1}}$).

(7) at later stages, the anisotropy relaxes though the fluctuations are still present (Figure \ref{f.3dpic}d, $t>44$ ${\Omega_{ci0}^{-1}}$).

(8) the fluctuations {at the edge} of the plasmoid are affected by the inhomogeneous plasma pressure growth inside the plasmoid and become strongly nonlinear at later stages (Figure \ref{f.3dpic}c).

{The combination of Cluster observations and PIC simulations leads to} the following interpretation. 
During multiple reconnection, {ions}, accelerated by reconnection, propagate mostly parallel to the magnetic field at the {edge} of the plasmoid forming between  {two} X-lines. As reconnection proceeds, the island contracts which
leads to additional acceleration of {ions} in the magnetic trap between the X-lines. As the temperature anisotropy reaches the firehose marginal stability threshold, the firehose instability {excites} magnetic field fluctuations, which are {confined} in the looped magnetic field. 
Further compression of magnetic field by ongoing reconnection may lead to the nonlinear evolution of {these }fluctuations and their transformation {in}to bent field line{s} {leading to the formation of a} thin boundary, similar to the one observed between the X-lines {after} {the temperature anisotropy was observed} \citep{alexandrova2016}. {Note }{that both active X-lines are observed,
thus reconnection does not cease with the development of instability, unlikely to what was {shown in simulations by}
\citet{drake2006, drake2010, burgess2016}}.
{In such periodic-boundary simulations}, the development of the firehose-induced fluctuations{,} together with the inhomogeneous pressure growth in the plasmoid center, constrains the plasmoid shrinking at later stages and initiate the phase of jet fronts {repulsion} with {consequent} decrease of the reconnection rate {at} the X-lines. 
Simulations with open {boundary conditions}{,} as well as observations covering larger temporal and {spatial} scale{s,} might be helpful in describing particular conditions for {different ways of the plasmoid-chain evolution}.

{It is} important to note that the performed stability analysis was based on the theoretical assumptions of a stable background magnetic field and {of} bi-Maxwellian particle distribution{s}, while none of these assumptions are rigorously valid in the observations. However, the consistency of the magnetic field fluctuations with the temperature anisotropy of plasma indicates that{,} despite {of these issues}, the firehose instability is {identified}. 

Previous studies of plasma stability in the context of temperature anisotropies in the solar wind \citep{hellinger2006, bale2009, matteini2013} and in magnetotail reconnection jets \citep{voros2011, hietala2015} performed statistical analysis of the average characteristics of magnetic field fluctuations and typical marginal stability thresholds. In the present study, we {analyzed in details} the development of the firehose instability and {of} the associated waves excitation, for the {specific} magnetic configuration of {multiple reconnection with a} plasmoid forming between two reconnection sites. The analysis showed that{,} differently to the average solar wind conditions, multiple reconnection results in highly inhomogeneous anisotropic plasma and nonlinear development of firehose-related fluctuations. Our simulations show that the anisotropy grows {far} beyond the marginal stability threshold before the firehose-related fluctuations arise. However, fluctuations are present even after plasma reaches the marginal stability. This {supports {the fact}} that the observations of relatively small growth rate, but quite intensive  fluctuations, indicate that the plasmoid in the magnetotail was detected {during} instability {decaying} and plasma isotropisation.

An important aspect is that the instability {may} affect the global energy conversion in the multiple reconnection configuration.
The magnetic energy conversion in the neighboring X-lines leads to particle acceleration, which in turn {leads} to the development of the firehose instability {at the edge of the plasmoid between the X-lines}. {A}s a consequence, energy of accelerated particles {is converted} back to magnetic field fluctuations. 
Understanding the ratio between the energy {converted} 
to plasma by reconnection and the energy withdrawn by the {firehose} instability
might help {to} better quantify the {impact of} current sheet disruption {to multiple reconnection for} space plasma dynamics 
and would be a valuable direction for a future research.

\section{Conclusions}

We studied {the} {dynamics of} {a} plasmoid between two active X-lines observed in situ by {the} Cluster spacecraft in the Earth's magnetotail current sheet.
{At th{e} edge} of the plasmoid having a scale of about $35$ ion inertial length{s} (about three Earth's radii),
{a} parallel temperature anisotropy of {ions} {due to} parallel and anti-parallel ion beams {was observed}.
The plasma conditions corresponded to {a} firehose marginal stability {state}, {as} also 
supported by the excitation of the whistler low-frequency branch waves  (at {about} half of the ion cyclotron frequency).
Reconstruction of the magnetotail observations by using PIC simulations
{allowed us to reproduce the evolution of} the firehose instability during multiple reconnection. {Simulations} support the scenario {in which} 
the looped magnetic field {of the plasmoid} between {two} reconnection sites {undergoes} fluctuations 
due to the {firehose instability caused by the} excess of parallel {ion} acceleration {at the edges} of the plasmoid.  
In such a way, some part of the plasma energy gained in reconnection 
might be converted back to magnetic field.
{The present study indicates that the firehose instability in multiple reconnection can play an important role for energy partition not only in the terrestrial magnetotail, but also in solar and astrophysics plasmas where multiple reconnection is expected to be ubiquitous.}

\acknowledgments
Authors acknowledge the Cluster Science Archive for use of the Cluster spacecraft data. The research was supported by the project of Sorbonne Universit\'{e}/Ecole Polytechnique, Convention 2800, and the LABEX PLAS@PAR project with the financial state aid managed by the Agence Nationale de la Recherche, as a part of the Programme 'Investissements d'Avenir' under the reference ANR-11-IDEX-0004-02. J.D. acknowledges support from NASA's Solar System Exploration Research Virtual Institute (SSERVI): Institute for Modeling Plasmas, Atmosphere, and Cosmic Dust (IMPACT), and the NASA High-End Computing (HEC) Program through the NASA Advanced Supercomputing (NAS) Division at Ames Research Center.
 
 
\bibliography{Bibliography}{}

\vspace{-5in}

\end{document}